\documentclass[sigconf,natbib=false]{acmart}

\AtBeginDocument{%
  }

\RequirePackage[
  datamodel=acmdatamodel,
  style=acmnumeric,
  ]{biblatex}
\makeatletter
\@addtoreset{subsubsection}{section}
\makeatother

\copyrightyear{2024}
\acmYear{2024}
\setcopyright{rightsretained}
\acmConference[CF '24 Companion]{Proceedings of the 21st ACM International Conference on Computing Frontiers Workshops and Special Sessions}{May 7--9, 2024}{Ischia, Italy}
\acmBooktitle{Proceedings of the 21st ACM International Conference on Computing Frontiers Workshops and Special Sessions (CF '24 Companion), May 7--9, 2024, Ischia, Italy}\acmDOI{10.1145/3637543.3652881}
\acmISBN{979-8-4007-0492-5/24/05}
\begin{document}



\title{A Gigabit, DMA-enhanced Open-Source Ethernet Controller for Mixed-Criticality Systems}


\author{Chaoqun Liang}
\affiliation{%
  \institution{University of Bologna}
  \city{Bologna}
  \country{Italy}}
\email{chaoqun.liang@unibo.it}

\author{Alessandro Ottaviano}
\authornote{Both authors contributed equally to this research.}
\author{Thomas Benz}
\authornotemark[1]

\affiliation{%
  \institution{ETH Z\"urich}
  \country{Switzerland}
}
\email{aottaviano@iis.ee.ethz.ch}
\email{tbenz@iis.ee.ethz.ch}
\author{Mattia Sinigaglia}
\affiliation{%
  \institution{University of Bologna}
  \city{Bologna}
  \country{Italy}
}\email{mattia.sinigaglia5@unibo.it}

\author{Luca Benini}
\affiliation{%
  \institution{ETH Z\"urich, University of Bologna}
  \country{Switzerland, Italy}
}
\email{luca.benini@unibo.it}

\author{Angelo Garofalo}
\affiliation{%
  \institution{ETH Z\"urich, University of Bologna}
  \country{Switzerland, Italy}
}\email{angelo.garofalo@unibo.it}

\author{Davide Rossi}
\affiliation{%
  \institution{University of Bologna}
  \city{Bologna}
  \country{Italy}
}\email{davide.rossi@unibo.it}

\renewcommand{\shortauthors}{Liang et al.}



\begin{abstract}
  The ongoing revolution in application domains targeting autonomous navigation, first and foremost automotive "zonalization", has increased the importance of certain off-chip communication interfaces, particularly Ethernet. The latter will play an essential role in next-generation vehicle architectures as the backbone connecting simultaneously and instantaneously the zonal/domain controllers.
  There is thereby an incumbent need to introduce a performant Ethernet controller in the open-source HW community, to be used as a proxy for architectural explorations and prototyping of mixed-criticality systems (MCSs).
  Driven by this trend, in this work, we propose a fully open-source, DMA-enhanced, technology-agnostic Gigabit Ethernet architecture that overcomes the limitations of existing open-source architectures, such as Lowrisc's Ethernet, often tied to FPGA implementation, performance-bound by sub-optimal design choices such as large memory buffers, and in general not mature enough to bridge the gap between academia and industry.
  Besides the area advantage, the proposed design increases packet transmission speed up to almost 3x compared to Lowrisc's and is validated through implementation and FPGA prototyping into two open-source, heterogeneous MCSs.
\end{abstract}



\maketitle

\section{Introduction}

Modern embedded applications and the Internet of Things (IoT) demand efficient communication protocols to meet their evolving needs.The rise of automotive "zonalization ~\cite{nxp2023zonal}" emphasizes Ethernet's crucial role in linking controllers within advanced vehicle architectures. Demanded sophisticated communication within drone applications necessitates the adoption efficient data transfer protocols like MAVLink~\cite{auterion2023mavlink} for component interaction. This trend necessitates a high-performance Ethernet controller for developing and prototyping mixed-criticality systems (MCSs)~\cite{5465960}.  Ethernet stands out in this context, offering reduced latency and higher data rates compared to traditional protocols, which is crucial for real-time processing and transmission of data in these applications~\cite{app13031278}. There is a particular emphasis ~\cite{830317} on developing 1Gbps Ethernet IPs because they provide significant improvements in terms of bandwidth and speed over their predecessors, enhancing the performance of networked systems.

Despite the advantages, most solutions in the Ethernet IP domain remain proprietary, restricting their application in academic research or developing open-source, heterogeneous architectures for embedded systems. An exception to this trend is the open-source Ethernet IP provided by LowRISC.~\cite{lowrisc}, representing a valuable community resource.

However, LowRISC's Ethernet IP is not without its limitations, which can impact its broader applicability:
\begin{itemize}
\item  The design is primarily tailored for FPGA implementation, lacking the necessary adaptability for ASIC integration. This limitation complicates its inclusion in heterogeneous System on Chip (SoC) designs.
\item  Data communication, both inbound and outbound, relies on internal buffers (one each for RX and TX). This reliance limits the maximum number of packets that can be sent in a single transaction, determined by the capacity of these buffers.
\item The architecture utilizes true dual-port memories for these buffers, components that are not universally supported across all technology nodes utilized in SoC prototyping.
\item Additionally, using a ping-pong mechanism for internal communication increases latency in end-to-end transactions, leading to performance degradation.
\end{itemize}

\begin{figure}[h]
  \centering
  \includegraphics[width=\linewidth]{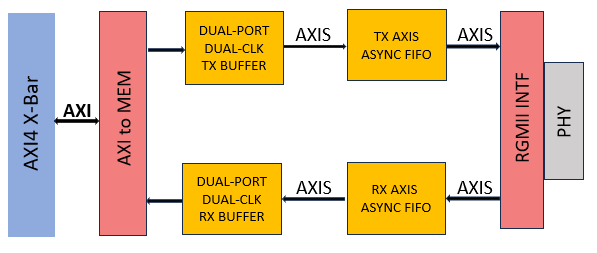}
  \caption{LowRisc Ethernet Design}
\end{figure}

To address these challenges, in this work we propose a buffer-less 1Gbps Ethernet IP. It integrates an autonomous iDMA engine that handles the communication between the AXI interface towards the system and the AXI Stream interface towards the Ethernet back-end, and seamlessly manages the Ethernet protocol without the need for internal buffering and ping-pong mechanisms. The design we propose is available open-source on GitHub. \footnote{\url{https://github.com/pulp-platform/pulp-ethernet.git}}

Specifically, the contributions of the work are as follows: 
\begin{itemize}
\item We enhance the capabilities of the existing open-source iDMA engine~\cite{benz2023highperformance} to handle the AXI Stream protocol, on which the Ethernet IP's front-end is based.
    \item We extend the baseline buffered Ethernet IP~\cite{lowrisc} by removing the dual-port memories (both for the transmission and reception channels) from the design and by integrating the extended iDMA.
    \item To evaluate our solution in terms of performance and area, we integrate the DMA-enhanced Ethernet controller into two open-source heterogeneous mixed-criticality architectures, namely Cheshire~\cite{ottaviano2023cheshire} and HULK-V~\cite{Valente_2023}, which target various embedded applications, including automotive, aerospace, and autonomous UAV navigation. To extract performance measurements on the communication latency, we implement the architectures on a Xilinx VCU118 FPGA, while to extract area estimations, we synthesize them targeting the GlobalFoundries' 22nm node.
    \item Finally, we benchmark and compare our solution against the baseline in terms of performance and area.
\end{itemize}




The evaluation results reveal that the proposed solution not only enhances flexibility and technological applicability but also demonstrates significant improvements in end-to-end communication latency and a notable reduction in the area footprint of the IP. 

\section{Architecture}

The proposed bufferless Ethernet IP is tailored for gigabit Ethernet applications. In this section, we describe its architecture, depicted in Fig. 2.It utilizes AXI-Stream-based processing for the sequential transfer of data between system components, optimizing point-to-point communication. The design is well-suited for high-throughput streaming applications with both efficiency and performance enhancement. The Media Access Controller (MAC) was integrated with reduced gigabit media-independent interface(RGMII) by LowRisc. Both RGMII transmitter and receiver are designed to support AXI Stream protocol. The use of AXI stream is motivated by its inherent protocol support for direct data flow management. This IP interfaces with an AXI4 interface, enhanced by an advanced iDMA (direct memory access) engine to efficiently manage data movement across the crossbar, connecting different components, memories, and peripherals. To bridge the iDMA, which operates with variable AXI Stream data width and clock depending on the system, to the RGMII interface with its fixed 125MHz clock and AXI stream width, the design incorporates CDC (clock domain crossing) FIFOs for clock domain synchronization, alongside AXI Stream down-sizer and up-sizer modules for ensuring data width compatibility. This design facilitates high-speed data transmissions while a unified register bus supports run-time configurability. In the following we provide architectural details of the Ethernet's sub-blocks.

\begin{figure}[h]
  \centering
  \includegraphics[width=\linewidth]{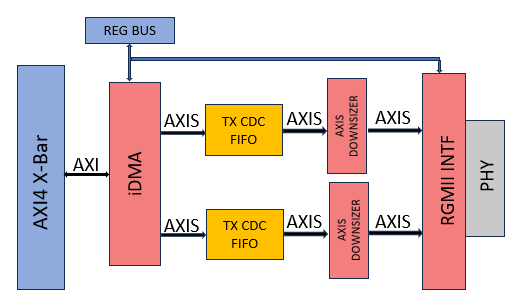}
  \caption{Bufferless Ethernet Design}
\end{figure}

\subsubsection {Ethernet Media Access Controller (MAC) with Reduced Gigabit Media Independent Interface (RGMII) ~\cite{xilinx2023xapp692}}

The original design proposed by LowRISC, as illustrated in Figure 1, features two dual-port buffers and two AXI Stream Asynchronous FIFOs. These buffers were utilized for buffering data, facilitating cross-clock domain operations, and adjusting data width up and down. Similarly, the two asynchronous FIFOs served the purpose of passing data through. However, due to significant area overhead and limitations imposed by technology dependencies, these components were subsequently removed, and the decision was made to expose the AXI Stream port directly to external connections.
The RGMII interface, to a cornerstone of the design, interfaces directly with AXI Stream data, facilitating high-speed Ethernet communications. It is specifically engineered to process data destined for the Physical Layer (PHY) at a 4-bit width, utilizing a 125 MHz clock with Double Data Rate (DDR) technology. This precise configuration is crucial for achieving the 1 Gbps Ethernet throughput,
as it allows for data transmission on both the rising and falling edges of the clock signal. By proficiently converting incoming AXI Stream data into the 4-bit DDR format required by the PHY interface, the RGMII block ensures a robust and efficient pathway for gigabit Ethernet data exchanges, highlighting its essential role in supporting advanced network connectivity.

\subsubsection{Direct Memory Access (DMA)}

The integration of iDMA emerges from the architectural advancements in the Ethernet interface, which now directly supports AXI Stream. iDMA's introduction is beneficial in this context, providing the flexibility to configure support for various protocols, including AXI and AXI Stream. This capability ensures efficient data transfer between Ethernet and system, optimizing network interactions by streamlining communications, enhancing throughput, and reducing latency. The direct support for AXI Stream by Ethernet IP, coupled with the configuration options of iDMA, achieves efficient and adaptable network exchanges.

iDMA features a request legalized for handling arbitrary one-dimensional transfer requests and an efficient transport layer for data movement, managing data flow between AXI and AXI Stream protocols, catering to both inbound and outbound Ethernet PHY data.
Besides, incorporating iDMA into the Ethernet IP transitions it from a passive subordinate to an active AXI manager, directly facilitating data exchanges and bypassing intermediary processes. This enhancement boosts system efficiency, enhancing throughput and reducing latency.

\subsubsection{Clock Domain Crossing (CDC) FIFOs}
The Ethernet controller, mandated to operate at 125 MHz to support 1 Gbps Ethernet, contrasts with the system's flexibility to function at various frequencies. This difference requires a bridge to ensure coherent data transfer between the differing clock domains of iDMA, which operates according to the system clock, and the Ethernet interface. To facilitate this seamless integration and maintain data integrity across these varied clock frequencies, TX and RX FIFOs are strategically deployed. These FIFOs act as critical intermediaries, buffering and synchronizing data between iDMA and Ethernet, thus enabling smooth data transfer and reinforcing the architecture's adaptability and efficiency in high-speed networking environments.

\subsubsection{Up/Down Sizers}
As the width of the  AXI Stream interface of the iDMA is wider than Ethernet controller, we integrated a down-sizer block, which is responsible for the data width adaptation, providing data compatibility between the two blocks. At the same time, we integrate an up-sizer block to expand the data bus width. 
The Up-Sizer and Down-Sizer are crucial in matching data packet sizes with the interface requirements, allowing for more efficient use of the available bandwidth. This alignment enables higher data transfer rates and maximizes throughput by ensuring that each communication cycle transmits the maximum amount of data supported by the receiving end.

\subsubsection{Configuration Registers}
The unified register file's runtime configurability allows for easy adjustments to system settings and parameters without the need for hardware modifications, streamlining the adaptation process to different operational conditions. Additionally, its standardized interface provides a common configuration point, reducing the complexity of integrating the system with various platforms and ensuring a straightforward deployment process. This simplification helps for rapid deployment and enhances compatibility across a wide range of technological ecosystems.

\section{Results}
To assess the benefits of our design, we integrate the two versions of the Ethernet IP, into two heterogeneous platforms: Cheshire, a small-size Linux-capable system~\cite{ottaviano2023cheshire} and HULK-V~\cite{Valente_2023}, a mixed-criticality platform targeting autonomous UAV navigation tasks.

We perform the evaluation and comparison in terms of area costs of the two solutions, and performance, measured as end-to-end latency of sending/receiving packets. 

To perform latency evaluation, we prototype the architectures on a Xilinx VCU118 FPGA; to perform area estimations, we synthesize the designs targeting the GlobalFoundries' 22nm FDX node in slow conditions and targeting a system frequency of 344 MHz.

This approach allowed us to meticulously gauge the IP’s operational effectiveness and its footprint, providing a comprehensive understanding of its applicability and scalability in high-speed networking environments.
In the following sections, we present the results of our experiments.

\subsubsection{Performance test}
In our latency performance evaluation, initial tests were conducted with a payload size of 2048 bytes. However, it was observed that the existing buffered Ethernet IP was limited by buffer capacity of 1536 bytes, preventing the initiation of transactions due to payload size constraints. Consequently, the testing scope was adjusted to encompass three distinct payload sizes:1024 bytes, 512 bytes, and 256 bytes. For each payload size, latency measurements were meticulously recorded across various transaction phases: configuration, preamble, payload, and CRC, with the conclusion of the CRC phase signaling the transaction’s end. The results highlighted in the corresponding figure reveal significant reductions in overall transaction cycles by the bufferless design, achieving savings of 62.97\%, 60.42\%, and 64.48\% for payload sizes of 256, 512, and 1024 bytes, respectively. These findings show the efficiency gains in reducing latency by adopting a bufferless Ethernet IP architecture.

Performance analysis is shown in Figure 3,

\begin{figure}[h]
  \centering
  \includegraphics[width=\linewidth]{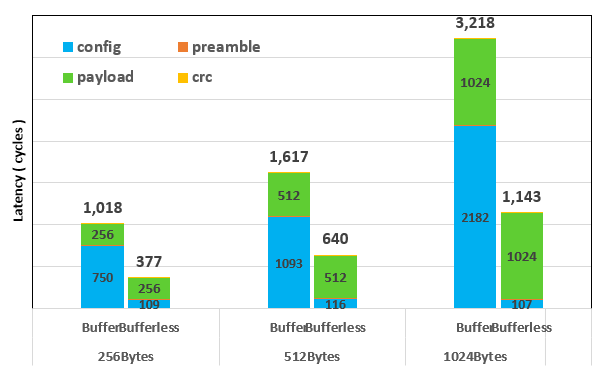}
  \caption{Performance Comparison of Buffer and Bufferless Ethernet Design}
\end{figure}

\subsubsection{Area test}
For the area efficiency evaluation, the design was synthesized within the Al-Saqr platform using GlobalFoundries 22nm technology. The analysis revealed that in the buffered design, the buffers occupied a substantial portion of the total system area, resulting in a total area footprint of
65,242.90 µm². Conversely, the bufferless design significantly reduced the total area to 7,754.04 µm², despite the inclusion of some area overheads attributed to the CDC FIFOs and DMA. This reduction translates to an impressive 88.12\% savings in area and highlights the substantial efficiency gains achieved through eliminating buffers in the design.
Figures 4 and 5 present the results from the area tests.

\begin{figure}[h]
  \centering
  \includegraphics[width=\linewidth]{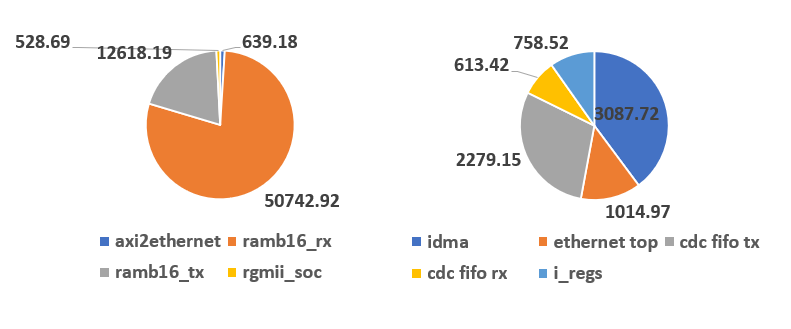}
  \caption{Buffer and Bufferless Area Info in GF22}
\end{figure}

\section{Conclusion}
We have presented a novel bufferless Ethernet IP architecture and its integration with direct data transfer mechanisms. We evaluated the system's latency performance and area efficiency using GlobalFoundries' 22nm node, employing a methodical approach to quantify improvements. Our findings demonstrate that by eliminating traditional buffers and incorporating an open-source DMA compatible with AXI Stream interfaces, our solution seamlessly integrates into any heterogeneous system featuring AXI. This approach reduces the area by an impressive 88.12\% compared to the baseline and significantly improves performance, evidenced by a marked improvement in latency. These advancements showcase our architecture's capacity to deliver efficient, high-speed networking capabilities while optimizing system resource utilization.

\begin{acks}
This work was supported in part through the TRISTAN (101095947) project that received funding from the Key Digital Technologies Joint Undertaking (KDT JU) under grant agreement nr. 101095947, and in part by the Spoke 1 on Future High-Performance-
Computing (HPC) of the Italian Research Center on High-Performance
Computing, Big Data and Quantum Computing (ICSC) that received funding
from the Ministry of University and Research (MUR) for the Mission 4–Next
Generation EU programme.
\end{acks}
\printbibliography

\appendix

\end{document}